\documentclass[aps,prmaterials,superscriptaddress,preprint,onecolumn]{revtex4-2}

\usepackage{multirow}
\usepackage{graphicx}
\usepackage{amsfonts,amsmath,amssymb,amsthm}
\usepackage{epstopdf}
\usepackage{upgreek,xspace}
\usepackage{chngcntr}
\usepackage{xr}
\usepackage{ulem}
\usepackage[colorlinks,allcolors=blue]{hyperref}
\usepackage[version=3]{mhchem} 

\newcommand {\rxx}{$\rho_\mathrm{xx}$}
\newcommand {\ryx}{$\rho_\mathrm{yx}$}

\begin{document}
\bibliographystyle{apsrev4-2}
\title{Anisotropic multiband magnetotransport in \ce{LaAg2Ge2} thin films}
\author{Mizuki~Ohno}
\affiliation{Department of Applied Physics and Materials Science, California Institute of Technology, Pasadena, California 91125, USA.}
\affiliation{Institute for Quantum Information and Matter, California Institute of Technology, Pasadena, California 91125, USA.}

\author{Reiley~Dorrian}
\affiliation{Department of Applied Physics and Materials Science, California Institute of Technology, Pasadena, California 91125, USA.}
\affiliation{Institute for Quantum Information and Matter, California Institute of Technology, Pasadena, California 91125, USA.}

\author{Veronica~Show}
\affiliation{Department of Applied Physics and Materials Science, California Institute of Technology, Pasadena, California 91125, USA.}
\affiliation{Institute for Quantum Information and Matter, California Institute of Technology, Pasadena, California 91125, USA.}

\author{Joseph~Falson}
\email{falson@caltech.edu}
\affiliation{Department of Applied Physics and Materials Science, California Institute of Technology, Pasadena, California 91125, USA.}
\affiliation{Institute for Quantum Information and Matter, California Institute of Technology, Pasadena, California 91125, USA.}

\begin{abstract}
ThCr$_2$Si$_2$-type intermetallics are layered conductors in which crystallographic anisotropy and multiband electronic states often give rise to characteristic magnetotransport phenomena.
Here, we report the molecular-beam epitaxy growth of \ce{LaAg2Ge2} thin films on MgO(001) and their magnetotransport properties.
The Hall effect and magnetoresistance are captured by an effective two-carrier description with a high-mobility electron band, yielding a positive magnetoresistance of 22.5\% at 9~T.
Angle-dependent magnetoresistance exhibits a dominant twofold anisotropy and additional reproducible dip/peak features at characteristic tilt angles that are nearly independent of field and temperature.
These results extend our understanding of the anisotropic electronic transport in thin-film germanides within the ThCr$_2$Si$_2$ family.
\end{abstract}

\maketitle
\section{Introduction}
ThCr$_2$Si$_2$-type (122) intermetallics, with general formula $AT_2X_2$, form one of the most extensively studied families of layered compounds~\cite{Ban1965_Crystal, Hoffmann1985_Making, Shatruk2019_ThCr2Si2a}.
In this structure, $T_2X_2$ layers alternate with $A$-atom planes along the $c$ axis, which often yields pronounced electronic anisotropy and a broad range of ground states, including magnetic order~\cite{HUAN1989_New, 1989_Chapter, Luo2012_Magnetism, Tan2018_Correlatinga}, unconventional and heavy-fermion superconductivity~\cite{Steglich1979_Superconductivity, Rotter2008_Superconductivity, Khim2021_Fieldinduced, Mydosh2020_Hidden, Nogaki2021_Topological}, and nontrivial transport responses~\cite{Peng2018_Crystal, Su2020_Magnetotransport, Malick2022_Large, Hooda2025_Electronic}.
For 122 compounds with $X=\mathrm{Si}$ or $\mathrm{Ge}$, band-structure calculations commonly predict multiple Fermi-surface sheets with different anisotropies, including quasi-two-dimensional and three-dimensional components~\cite{Zwicknagl2007_Kondoa, DucDung2008_Haasb, Yasui2011_Electronica, Ciesielski2018_Electronic}.
Because orbital motion in a magnetic field depends sensitively on field orientation, angle-dependent magnetoresistance can provide a stringent experimental probe of anisotropic and multiband transport to help understand the underlying electronic structure of these compounds.

\begin{figure}[b]
    \includegraphics[width=0.8\columnwidth]{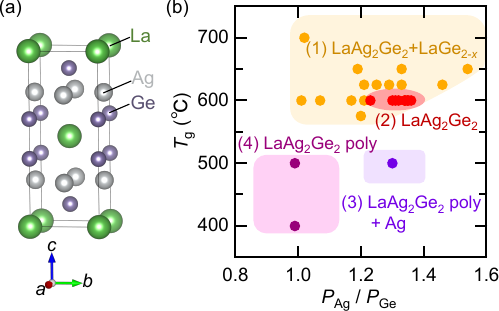}
    \caption{
			(a) Crystal structure of \ce{LaAg2Ge2} drawn with VESTA~\cite{Momma2011}.
            (b) Growth phase diagram of \ce{LaAg2Ge2} films on MgO(001) substrates.
            The diagram is mapped as a function of the growth temperature $T_{\mathrm{g}}$ and the flux ratio $P_{\mathrm{Ag}}/P_{\mathrm{Ge}}$ and is classified into four regions: (1) coexistence of \ce{LaAg2Ge2} and LaGe$_{2-x}$ phases, (2) single-phase \ce{LaAg2Ge2}, (3) coexistence of polycrystalline \ce{LaAg2Ge2} and Ag, and (4) polycrystalline \ce{LaAg2Ge2}.
			}\label{fig:1}
\end{figure}

Epitaxial thin films are well suited for such measurements because a single crystallographic orientation allows the magnetic-field direction to be swept continuously with respect to the principal axes.
However, thin-film studies of Si/Ge-based 122 compounds have thus far been limited to heavy fermion compounds such as Ce$T_2$Ge$_2$ ($T=\mathrm{Fe}$, $\mathrm{Ni}$, or $\mathrm{Cu}$)~\cite{Holter1986_Thinfilm, Schmied1998_Preparation, Li2011_Growtha} and \ce{YbRh2Si2}~\cite{Prochaska2020_Singulara, Bakali2022_Knudsen}, for which only zero-field resistivity has been reported.
Systematic magnetotransport measurements that exploit field-angle control in epitaxial 122 films therefore remain scarce.
\ce{LaAg2Ge2} crystallizes in the ThCr$_2$Si$_2$-type structure (space group $I4/mmm$)~\cite{Benndorf2024_Multinuclear}, with alternating La and Ag$_2$Ge$_2$ layers along the $c$ axis [Fig.~\ref{fig:1}(a)].
Because La$^{3+}$ has no $4f$ electrons, magnetic scattering and Kondo effects are absent, making \ce{LaAg2Ge2} an ideal system for isolating intrinsic charge-carrier transport.
To date, investigations have been limited to bulk polycrystalline samples~\cite{Thamizhavel2007_Anisotropica, Benndorf2024_Multinuclear}; specific-heat measurements established a nonmagnetic ground state, but electronic-transport properties have not been reported.

In this work, we report the synthesis of \ce{LaAg2Ge2} thin films grown on MgO(001) substrates and their unique magnetotransport properties. By utilizing the single-orientation nature of the films, we probe the intrinsic transport anisotropy and multiband features via angle-dependent magnetoresistance ratio (ADMR) and Hall measurements. Our study demonstrates the feasibility of growing high-quality 122-type germanide films and provides a baseline for understanding the charge-carrier transport in this material class.

\section{Experimental Details}
\ce{LaAg2Ge2} thin films were synthesized using a molecular beam epitaxy (MBE) system with a base pressure of $\sim 10^{-10}$~mbar on MgO(001) substrates that were laser-annealed to optimize surface quality~\cite{llanos_supercell:2024,llanos:2024,dorrian:2025}.
The in-plane lattice mismatch between \ce{LaAg2Ge2}~\cite{Benndorf2024_Multinuclear} and MgO is 2.9\%.
The growth temperature and beam equivalent pressures, measured by an ionization gauge, were tuned to obtain single-phase films.
The film thickness was typically 32~nm, corresponding to a growth rate of approximately 0.045~\AA{}/s.
To protect the films from degradation in air, all samples not examined by atomic force microscopy were capped \textit{in situ} with an amorphous Ge layer ($\approx$5~nm) deposited at room temperature.

Structural properties of the films were characterized by X-ray diffraction (XRD) (SmartLab, Rigaku). 
Surface morphology of the uncapped film was examined by atomic force microscopy in a \ce{N2}-filled glovebox.
Electrical transport measurements above 1.8~K were carried out in Quantum Design DynaCool Physical Properties Measurement Systems (PPMS) equipped with 9~T and 14~T superconducting magnets.
The film was cut into a rectangular piece ($2\times5$~mm) and contacted with Al wires at six points for simultaneous four-terminal measurements of the longitudinal resistivity (\rxx{}) and the Hall resistivity (\ryx{}).
For standard magnetotransport, the magnetic field was applied perpendicular to the film plane; the longitudinal and Hall components were separated by symmetrization and antisymmetrization with respect to the field direction.
For ADMR measurements, the magnetic field was rotated in the $ac$ plane from the out-of-plane direction ($\theta=0^\circ$, $B\parallel c$) to the in-plane direction ($\theta=90^\circ$, $B\parallel a$) using a sample rotator, while keeping $I \parallel a$.
    
\section{Results and discussion}
The growth temperature and the flux ratio between Ag and Ge were systematically varied while maintaining a constant La flux of $1.0\times10^{-8}$~mbar, as summarized in Fig.~\ref{fig:1}(b).
Under optimized conditions, single-phase \ce{LaAg2Ge2} is confirmed by XRD $\theta$--$2\theta$ scans without impurity peaks (region~2).
In contrast, impurity phases appear when the growth conditions deviate from this narrow window.
At temperatures above $600\,^{\circ}\mathrm{C}$ and for smaller flux ratios $P_{\mathrm{Ag}}/P_{\mathrm{Ge}} \lesssim 1.2$, a Ge-rich LaGe$_{2-x}$ phase emerges, consistent with Ag deficiency caused by enhanced re-evaporation of Ag at elevated temperatures (region~1).
For growth temperatures below $500\,^{\circ}\mathrm{C}$, polycrystalline \ce{LaAg2Ge2} peaks are observed (regions~3 and~4).
This behavior is attributed to insufficient surface diffusion at low temperatures, which suppresses epitaxial ordering and leads to polycrystalline domains.
When $P_{\mathrm{Ag}}/P_{\mathrm{Ge}}$ is increased beyond the optimal range, Ag impurity peaks appear (region~3).
Representative XRD $\theta$--$2\theta$ scans for regions~1--4 are provided in Fig.~S1 of the Supplemental Material~\cite{Supplement}.
 
\begin{figure*}[b]
    \includegraphics[width=1\textwidth]{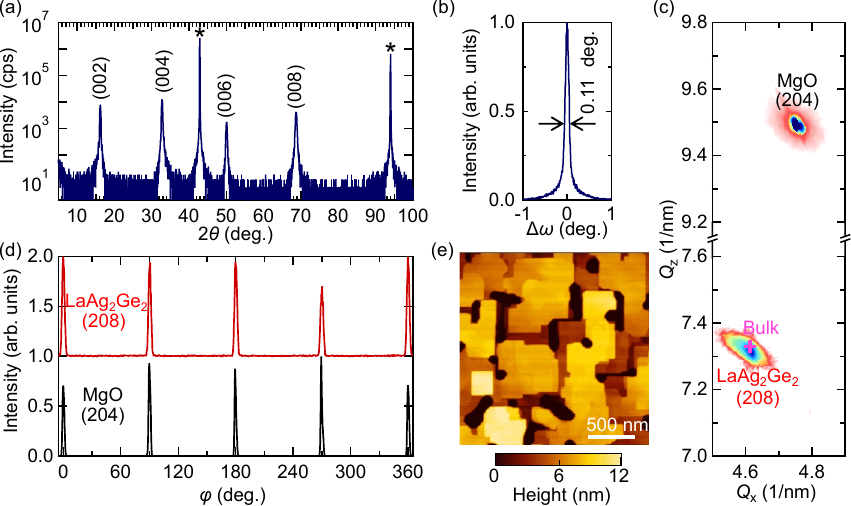}
    \caption{
			  (a) XRD $\theta$--2$\theta$ scan of a \ce{LaAg2Ge2} film grown on a MgO(001) substrate.
            MgO substrate peaks are marked with an asterisk.
			  (b) Rocking curve around the (002) \ce{LaAg2Ge2} film peak.
			(c) Reciprocal space map around the (208) \ce{LaAg2Ge2} film and (204) MgO substrate peaks.
            The pink cross indicates the peak position expected from bulk polycrystalline lattice parameters~\cite{Benndorf2024_Multinuclear}.
            (d) In-plane azimuthal ($\varphi$) scans around \ce{LaAg2Ge2} (208) (top) and MgO (204) (bottom) peaks.
            (e) Surface morphology of the film, taken by atomic force microscopy.
			}\label{fig:2}
\end{figure*}

Figure~\ref{fig:2} shows the structural characterization of a single-phase \ce{LaAg2Ge2} film.
The XRD ${\theta}$--${2\theta}$ scan in Fig.~\ref{fig:2}(a) reveals sharp reflections from the \{001\} \ce{LaAg2Ge2} lattice planes without any impurity phases.
The out-of-plane lattice parameter along the $c$ axis is determined to be $c = 10.920$~\AA.
The full width at half maximum (FWHM) of the rocking curves around the peak of (002) (Fig.~\ref{fig:2}(b)) is about 0.11$^\circ$, indicating the high crystalline quality of the film.
The reciprocal-space map in Fig.~\ref{fig:2}(c) yields an in-plane lattice parameter $a=4.33$~\AA, slightly smaller than the bulk value $a=4.3354$~\AA~\cite{Benndorf2024_Multinuclear}, while the out-of-plane parameter slightly exceeds the bulk value $c=10.914$~\AA.
This tetragonal distortion is consistent with a small compressive in-plane strain imposed by the MgO substrate ($a_{\mathrm{MgO}}=4.212$~\AA).
The azimuthal ($\varphi$) scans of the \ce{LaAg2Ge2} (208) and MgO (204) reflections in Fig.~\ref{fig:2}(d) display fourfold symmetry at the same azimuthal positions, confirming single-domain growth with a well-defined in-plane orientation relationship.
The atomic force microscopy image in Fig.~\ref{fig:2}(e) shows a terraced morphology with facets aligned along orthogonal directions, reflecting the tetragonal symmetry of \ce{LaAg2Ge2}.

\begin{figure}[b]
\begin{center}
\includegraphics*[width=0.8\columnwidth]{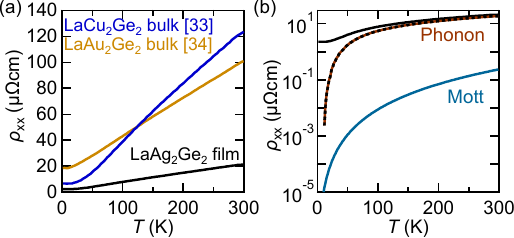}
\caption{
(a) Temperature dependence of the longitudinal resistivity {\rxx}($T$) for \ce{LaAg2Ge2} thin film (present study), compared with previously reported bulk single-crystalline samples of \ce{LaCu2Ge2}~\cite{Mendpara2015_Magnetica} and \ce{LaAu2Ge2}~\cite{Joshi2010_Magnetocrystalline}.
(b) Bloch--Gr\"uneisen--Mott (BGM) fitting of the $\rho_\mathrm{xx}(T)$ data for the \ce{LaAg2Ge2} film.
The solid black line shows the experimental $\rho_\mathrm{xx}(T)$ data, while the dashed line represents the temperature-dependent resistivity $\rho_\mathrm{xx}(T)-\rho_0$ obtained by subtracting the residual resistivity $\rho_0$.
The individual phonon (orange) and Mott (blue) contributions obtained from the BGM model are also shown.
}
\label{fig:3}
\end{center}
\end{figure}

We next discuss the temperature dependence of the longitudinal resistivity \rxx{} of our films. Figure~\ref{fig:3}(a) shows the longitudinal resistivity $\rho_\mathrm{xx}(T)$ of the \ce{LaAg2Ge2} film together with previously reported data for single-crystalline bulk samples of \ce{LaCu2Ge2}~\cite{Mendpara2015_Magnetica} and \ce{LaAu2Ge2}~\cite{Joshi2010_Magnetocrystalline}.
Within the measured temperature range and measurement geometry, the \ce{LaAg2Ge2} film exhibits a lower $\rho_\mathrm{xx}$ than these literature data.
To gain further insight into the scattering mechanisms governing the resistivity of \ce{LaAg2Ge2}, we analyze the $\rho_\mathrm{xx}(T)$ data using the Bloch--Gr\"{u}neisen--Mott (BGM) model~\cite{Ziman2001_Electrons, Bid2006_Temperature, Joshi2010_Magnetocrystalline, Ciesielski2018_Electronic},
\begin{equation}
\rho_\mathrm{xx}(T)=\rho_0
+4\Theta_{\mathrm{D}} R\left(\frac{T}{\Theta_{\mathrm{D}}}\right)^5
\int_0^{\Theta_{\mathrm{D}}/T}
\frac{x^5\,dx}{(e^x-1)(1-e^{-x})}
-KT^3,
\label{eq:BGM_LaTE2Ge2_1}
\end{equation}
where $\rho_{0}$ is the residual resistivity.
The second term is the Bloch--Gr\"uneisen expression for electron--phonon scattering with an effective phonon cutoff parameter $\Theta_{\mathrm D}$ (commonly used as a transport proxy for the Debye temperature), and $R$ is the corresponding prefactor.
The last term represents the Mott $s$--$d$ interband-scattering contribution, written as $\rho_{sd}(T)=-KT^3$; in the fitting procedure, $K$ is allowed to take either sign to accommodate material-dependent curvature in $\rho(T)$.
As shown in Fig.~\ref{fig:3}(b), $\rho_\mathrm{xx}(T)$ of \ce{LaAg2Ge2} is well reproduced by the BGM model.
The fitted parameters are summarized in Table~\ref{tab:transport} and are comparable to those reported for related intermetallic compounds~\cite{Ciesielski2018_Electronic}.
To assess the importance of the Mott term, it is instructive to compare the temperature-dependent contributions to $\rho(T)$.
For \ce{LaAg2Ge2}, the extracted $K$ yields a Mott contribution that remains small compared to the phonon term over the entire measured temperature range, indicating that the resistivity is dominated by electron--phonon scattering up to room temperature.

\begin{table*}[b]
\caption{Transport and phonon parameters extracted from Bloch--Gr\"uneisen--Mott (BGM) analysis for bulk \ce{LaCu2Ge2}~\cite{Mendpara2015_Magnetica} and \ce{LaAu2Ge2}~\cite{Joshi2010_Magnetocrystalline} and a thin film \ce{LaAg2Ge2} sample.}
\label{tab:transport}
\begin{ruledtabular}
\begin{tabular}{l|ccccc}
Compound & RRR & $\rho_0$ ($\mu\Omega$\,cm) & $\Theta_\mathrm{D}$ (K) & $R$ ($\mu\Omega$\,cm\,K$^{-1}$) & $K$ ($\mu\Omega$\,cm\,K$^{-3}$) \\
\hline
\ce{LaCu2Ge2} bulk & 18.7 & 7.0  & 222 & 0.42 & $1.7\times10^{-7}$ \\
\ce{LaAg2Ge2} film & 9.3  & 2.3  & 181 & 0.07 & $8.8\times10^{-9}$ \\
\ce{LaAu2Ge2} bulk & 5.6  & 19.0 & 126 & 0.27 & $-5.8\times10^{-8}$ \\
\end{tabular}
\end{ruledtabular}
\end{table*}

A systematic trend is observed in the fitted $\Theta_{\mathrm D}$ values, increasing from \ce{LaAu2Ge2} to \ce{LaAg2Ge2} and \ce{LaCu2Ge2} (Table~\ref{tab:transport}).
While $\Theta_{\mathrm D}$ from BGM fitting should be interpreted as an effective phonon cutoff rather than a thermodynamic Debye temperature, the trend is qualitatively consistent with variations in atomic mass and bonding strength across the series~\cite{Stewart1983_Measurementa}.
Within the BGM framework, the high-temperature resistivity is largely governed by the phonon contribution, which depends on both the characteristic phonon scale ($\Theta_{\mathrm D}$) and the overall coupling strength captured by the prefactor $R$.
Consistent with this picture, \ce{LaCu2Ge2} exhibits both the largest $\Theta_{\mathrm D}$ and the largest $R$ in Table~\ref{tab:transport}, resulting in a substantially larger $\rho(300\,\mathrm{K})$ than \ce{LaAg2Ge2}.
Accordingly, although the \ce{LaAg2Ge2} film shows a small residual resistivity, its comparatively smaller phonon-scattering contribution at high temperature reduces $\rho(300\,\mathrm{K})$ and thus yields a smaller RRR than \ce{LaCu2Ge2}.
We note that RRR is also sensitive to sample-dependent elastic scattering through $\rho_0$; therefore, any apparent correlation between $\Theta_{\mathrm D}$ and RRR should be regarded as qualitative rather than strictly causal.
This interpretation is consistent with trends reported for other nonmagnetic \ce{La$M_2$Ge2} compounds~\cite{Ciesielski2018_Electronic}.

\begin{figure}[b]
		\begin{center}
			\includegraphics*[width=0.8\columnwidth]{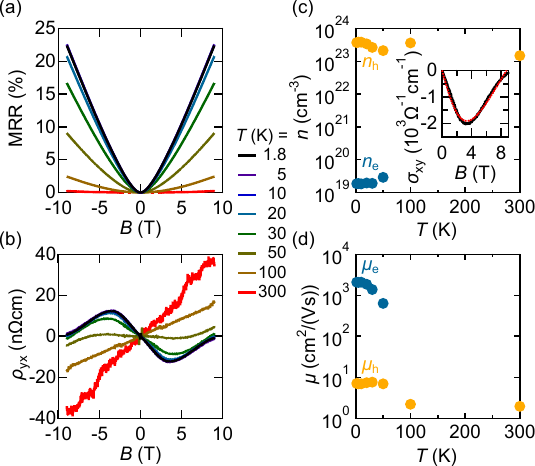}
			\caption{
            Magnetic field (\textit{B}) dependence of (a) magnetoresistance ratio (MRR $\equiv \rho_{\mathrm{xx}}(B)/\rho_{\mathrm{xx}}(0)-1$) and (b) Hall resistivity ({\ryx}) under an out-of-plane magnetic field at a series of temperatures.
            Temperature dependence of (c) carrier density and (d) mobility as estimated by fits to the two-carrier Drude model.
            The values at 100 and 300~K are obtained from single-carrier linear fitting.
			}\label{fig:4}
		\end{center}
\end{figure}

Figure~\ref{fig:4} summarizes the magnetotransport properties of the \ce{LaAg2Ge2} film.
Figures~\ref{fig:4}(a) and \ref{fig:4}(b) present the magnetoresistance ratio (MRR $\equiv \rho_{\mathrm{xx}}(B)/\rho_{\mathrm{xx}}(0)-1$) and Hall resistivity (\ryx), respectively, measured at various temperatures.
The \ce{LaAg2Ge2} film exhibits a positive magnetoresistance, the magnitude of which increases with decreasing temperature, reaching up to 22.5\% at 1.8~K and 9~T.
This behavior is consistent with conventional orbital magnetoresistance governed by carrier mobility.
At low temperatures, \ryx\ exhibits a pronounced nonlinear field dependence, indicating the coexistence of multiple types of charge carriers.
Moreover, both the Hall resistivity and its slope change sign with increasing magnetic field, providing clear evidence for the presence of electrons and holes in \ce{LaAg2Ge2}.
At higher temperatures, the $\rho_\mathrm{yx}$--$B$ curves tend to become linear, particularly above 100~K, suggesting that a single type of carrier dominates transport at elevated temperatures.

To quantify the carrier properties, the Hall conductivity $\sigma_\mathrm{xy}$ was analyzed using a two-carrier Drude model,
\begin{equation}
\begin{split}
\sigma_\mathrm{xy}(B)
&=
\frac{\rho_\mathrm{yx}(B)}{\rho_\mathrm{xx}^2(B) + \rho_\mathrm{yx}^2(B)}\\
&=
\left[
\frac{n_\mathrm{h} \mu_\mathrm{h}^2}{1 + (\mu_\mathrm{h} B)^2}
-
\frac{n_\mathrm{e} \mu_\mathrm{e}^2}{1 + (\mu_\mathrm{e} B)^2}
\right] eB,
\label{eq:two_carrier}
\end{split}
\end{equation}
where $n_\mathrm{h}$ ($n_\mathrm{e}$) and $\mu_\mathrm{h}$ ($\mu_\mathrm{e}$) denote the carrier densities and mobilities of holes (electrons), respectively.
Figures~\ref{fig:4}(c) and \ref{fig:4}(d) show the temperature dependence of the carrier density and mobility extracted from the two-carrier fits.
At 1.8~K, the fitting yields carrier densities of $n_\mathrm{h} = 3.7 \times 10^{23}~\mathrm{cm^{-3}}$ for holes and $n_\mathrm{e} = 1.9 \times 10^{19}~\mathrm{cm^{-3}}$ for electrons, with corresponding mobilities of $\mu_\mathrm{h} = 7~\mathrm{cm^{2}\,V^{-1}\,s^{-1}}$ and $\mu_\mathrm{e} = 2100~\mathrm{cm^{2}\,V^{-1}\,s^{-1}}$.
Electrons exhibit much higher mobility than holes and their density is more than three orders of magnitude smaller.
Holes dominate the total carrier density and show only weak temperature dependence.

With increasing temperature, the mobility of the minority electron decreases rapidly, reducing its weight in the total conductivity.
Accordingly, the Hall response becomes nearly linear above 100~K and can be described by an effective single-carrier picture.
While the two-carrier Drude parameters should be regarded as effective quantities, the analysis robustly indicates the coexistence of opposite-sign carriers and, in particular, a high-mobility minority electron contribution at low temperatures.
Such a high-mobility component enhances the orbital magnetoresistance and is consistent with the observed positive magnetoresistance.

\begin{figure}[b]
  \begin{center}
    \includegraphics*[width=0.65\columnwidth]{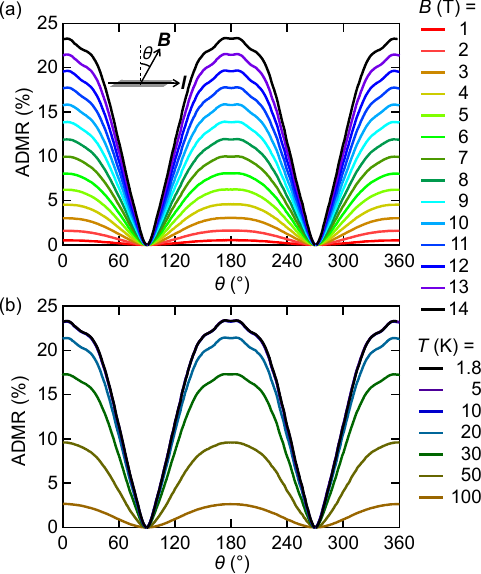}
    \caption{
    Angle-dependent magnetoresistance ratio (ADMR), defined as $\mathrm{ADMR}\equiv \rho_{\mathrm{xx}}(\theta)/\rho_{\mathrm{xx}}(90^{\circ})-1$, measured with $I \parallel a$ while rotating $B$ in the $ac$ plane.
    (a)~Field dependence at $T=1.8$~K. (b)~Temperature dependence at $B=14$~T.
    The inset in (a) illustrates the measurement geometry: the field angle $\theta$ is measured from the film normal ($[001]$), so that $\theta=0^\circ$ corresponds to $B \parallel c$ and $\theta=90^\circ$ to $B \parallel a$. 
    }
    \label{fig:5}
  \end{center}
\end{figure}

We measured the field-angle dependence of $\rho_\mathrm{xx}$ to characterize anisotropic magnetotransport in the epitaxial \ce{LaAg2Ge2} film (Fig.~\ref{fig:5}).
The ADMR was obtained by rotating the magnetic field from the out-of-plane direction ($\theta=0^\circ$, $B\parallel c$) toward the in-plane direction ($\theta=90^\circ$, $B\parallel a$), with the transport current applied along the $a$ axis.
Figure~\ref{fig:5}(a) shows the ADMR at $T=1.8$~K for several magnetic fields.
A pronounced twofold angular modulation is observed, with minima at $\theta=90^\circ$ and $270^\circ$ (field in the film plane).
At $B=14$~T the ADMR reaches 24\%, and its amplitude decreases monotonically with decreasing field while the overall angular profile is largely preserved.
Figure~\ref{fig:5}(b) displays the temperature evolution at a fixed field of 14~T: the ADMR is gradually suppressed upon warming, consistent with reduced carrier mobility, whereas the twofold symmetry persists over the entire temperature range.

In addition to the dominant twofold component, fine structure appears near $\theta \approx 0^{\circ}$ and $180^\circ$ in the low-temperature, high-field regime.
Although experimental factors such as a small angular misalignment or a weak Hall admixture could, in principle, produce a fixed-angle artifact, the dip/peak features are reproducible and $\rho_{\mathrm{xx}}(\theta)$ exhibits $\theta \rightarrow \theta+180^\circ$ symmetry, supporting an intrinsic origin.
Specifically, dip features are observed near $\theta \simeq 0^\circ$ and $20^\circ$, accompanied by peak features near $\theta \simeq 7^\circ$ and $31^\circ$.
These dip/peak angles remain essentially unchanged over the investigated field and temperature ranges, while their amplitudes grow with increasing field and decreasing temperature.
The nearly field- and temperature-independent angular positions suggest that these features are set primarily by Fermi-surface geometry, as expected for an orbital magnetotransport effect governed by the Fermi-surface anisotropy.
A plausible interpretation is a multiband ADMR response arising from the coexistence of Fermi-surface sheets with different anisotropies (e.g., three-dimensional and quasi-two-dimensional components), which can generate extrema at characteristic tilt angles set by the Fermi-surface geometry~\cite{Ali2016_Butterfly, Yuan2016_Large, Cheng2025_Anomalousc, Zhao2019_Extremely, Wang2019_Angledependentc}.
Possible alternative mechanisms, such as Yamaji oscillations~\cite{ Yamaji1989_Angle} andLebed magic-angle resonances~\cite{Lebed1989_Theory, Osada1992_Resonance, Maki1992_Lebeda}, are discussed further in the Supplemental Material (Fig.~S3)~\cite{Supplement}.

\section{Conclusion}
We have grown \ce{LaAg2Ge2} on MgO (001) thin films by MBE and characterized their structural and magnetotransport properties.
The films exhibit metallic conduction and a positive transverse magnetoresistance reaching 22.5\% at 1.8~K and $B=9$~T.
The low-temperature Hall nonlinearity is well described by a two-carrier model, revealing the coexistence of electron and hole contributions with strongly different mobilities, including a high-mobility minority electron.
ADMR measurements show a robust twofold anisotropy and, at low temperatures and high fields, additional dip/peak structures at specific field angles whose positions are insensitive to field strength and temperature, consistent with a Fermi-surface-related anisotropic magnetotransport response in this 122 germanide.
These results establish Ag--Ge-based 122 epitaxial films as a practical platform for controlled-orientation magnetotransport studies and motivate extensions to magnetically and electronically active 122 analogues.

\section{Acknowledgments}
We acknowledge funding provided by the Institute for Quantum Information and Matter, a NSF Physics Frontiers Center (NSF Grant PHY-2317110) and the Gordon and Betty Moore Foundation’s EPiQS Initiative (Grant number GBMF10638). We acknowledge the support of JSPS Overseas Research  Fellowships. This material is based upon work supported by the National Science Foundation Graduate Research Fellowship Program under Grant No. 2139433 (RD, VS).  Any opinions, findings,and conclusions or recommendations expressed in this material are those of the author(s) and do not necessarily reflect the views of the National Science Foundation.

\section*{References}
\bibliography{bib_arXiv}
\clearpage

\begin{center}
\Large\bfseries Supplementary Materials
\end{center}

\setcounter{section}{0}
\setcounter{figure}{0}

\renewcommand{\thesection}{S\arabic{section}}
\renewcommand{\thefigure}{S\arabic{figure}}
\section{Optimization of growth conditions}
\begin{figure}[h]
    \includegraphics[width=1\textwidth]{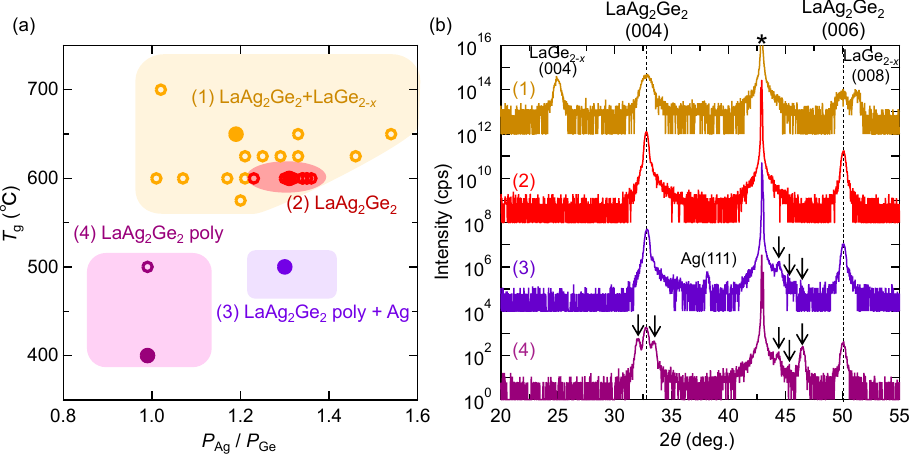}
    \caption{
            (a) Growth phase diagram of \ce{LaAg2Ge2} films on MgO(001) substrates.
            The diagram is mapped as a function of the growth temperature $T_{\mathrm{g}}$ and the flux ratio $P_{\mathrm{Ag}}/P_{\mathrm{Ge}}$ and is classified into four regions: (1) coexistence of \ce{LaAg2Ge2} and LaGe$_{2-x}$ phases, (2) single-phase \ce{LaAg2Ge2}, (3) coexistence of polycrystalline \ce{LaAg2Ge2} and Ag, and (4) polycrystalline \ce{LaAg2Ge2}. The filled circles indicate the growth conditions for which the XRD data shown in (b) were collected.
            (b) Representative XRD $\theta$--$2\theta$ scans corresponding to these four regions. Black arrows indicate polycrystalline diffraction peaks of \ce{LaAg2Ge2} other than the (004) and (006) reflections.
			}
    \label{fig:S1}
\end{figure}

We optimized the growth conditions by supplying La, Ag, and Ge fluxes onto MgO(001) substrates.
The growth temperature and the flux ratio between Ag and Ge were systematically varied while maintaining a constant La flux of $1.0\times10^{-8}$~mbar, as summarized in Fig.~\ref{fig:S1}.
When the flux ratio and substrate temperature are optimized ($P_{\mathrm{Ag}} = 1.2\times10^{-8}$~mbar, $P_{\mathrm{Ge}} = 0.9\times10^{-8}$~mbar, and $T_{\mathrm{g}} = 600\,^{\circ}\mathrm{C}$), a single-phase \ce{LaAg2Ge2} film is confirmed by XRD $\theta$--$2\theta$ scans without any impurity peaks (region~2).

In contrast, impurity phases appear when the growth conditions deviate from this narrow window.
At temperatures above $600\,^{\circ}\mathrm{C}$ and for smaller flux ratios $P_{\mathrm{Ag}}/P_{\mathrm{Ge}} \lesssim 1.2$, a Ge-rich LaGe$_{2-x}$ phase emerges, indicating Ag deficiency caused by enhanced re-evaporation of Ag at elevated temperatures (region~1).
For growth temperatures below $500\,^{\circ}\mathrm{C}$, polycrystalline \ce{LaAg2Ge2} peaks are observed (regions~3 and~4).
This behavior is attributed to insufficient surface diffusion of adatoms at low temperatures, which suppresses epitaxial ordering and leads to the formation of polycrystalline domains.
When the $P_{\mathrm{Ag}}/P_{\mathrm{Ge}}$ ratio is increased beyond the optimal range, Ag impurity peaks appear, corresponding to region~3.
 
\clearpage
\section{Structural characterization}

\begin{figure}[h]
    \centering
    \includegraphics[width=0.8\linewidth]{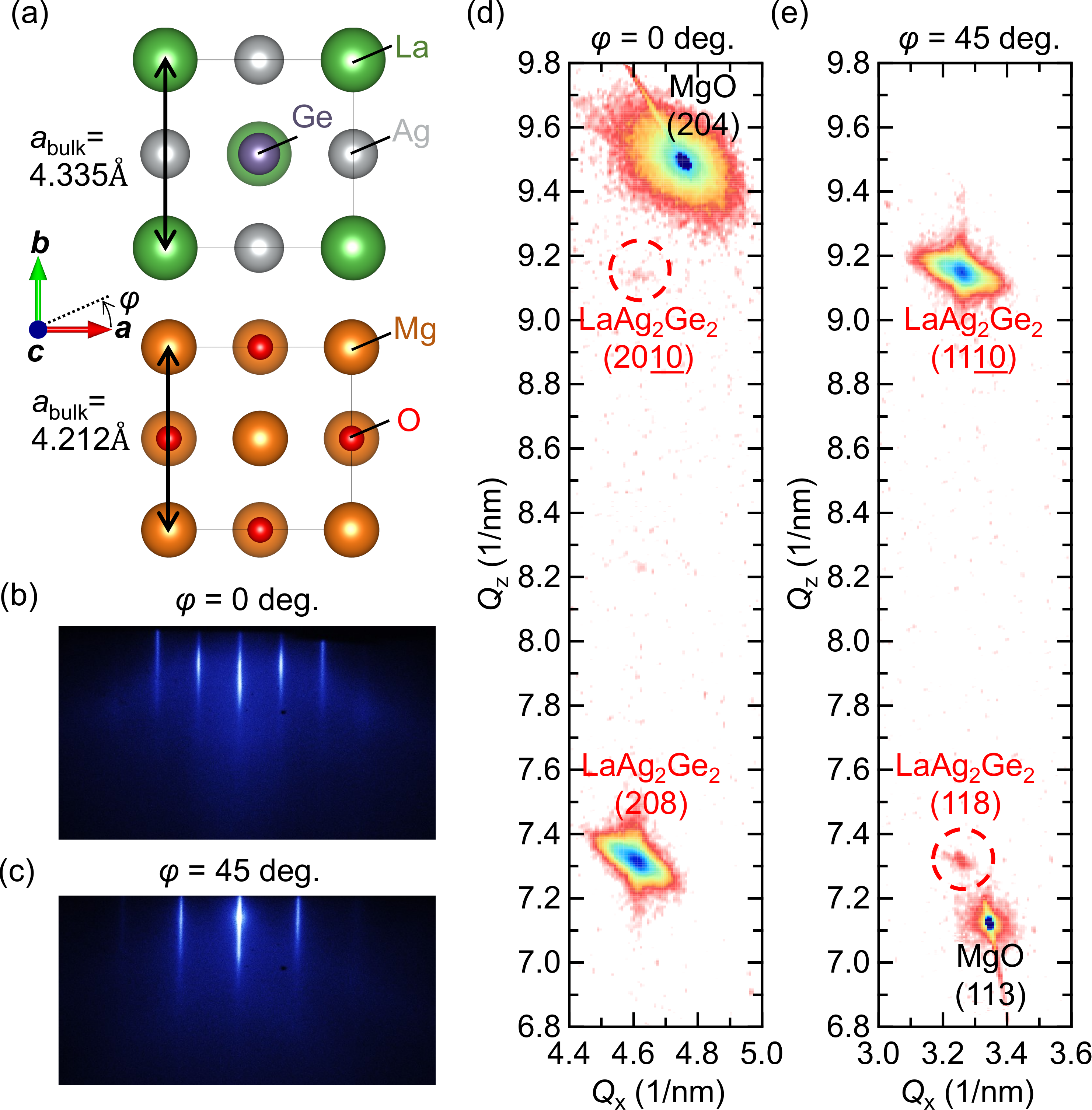}
    \caption{
    (a) Crystal structures of the \ce{LaAg2Ge2} film and the MgO substrate viewed along the out-of-plane direction.
    The angle $\varphi$ is defined as the angle between the incident electron beam and the crystallographic $a$ axis, as used in panels (b)--(e).
    \textit{In situ} RHEED patterns acquired after growth for (b) $\varphi = 0^\circ$ and (c) $\varphi = 45^\circ$.
    (d) Extended reciprocal-space mapping measured along the $h = 2$ rod at $\varphi = 0^\circ$ and
    (e) along the $h = k = 1$ rod at $\varphi = 45^\circ$.
    The relative intensities of the four \ce{LaAg2Ge2} diffraction peaks are consistent with those calculated from the bulk crystal structure~\cite{Benndorf2024_Multinuclear}.
    }
    \label{fig:S2}
\end{figure}

Figure~\ref{fig:S2} provides additional structural characterization of the \ce{LaAg2Ge2} film.
The in-plane lattice mismatch between \ce{LaAg2Ge2}~\cite{Benndorf2024_Multinuclear} and MgO is 2.9\%, as illustrated in Fig.~\ref{fig:S2}(a).
The \textit{in situ} reflection high-energy electron diffraction (RHEED) patterns acquired after growth are shown in Figs.~\ref{fig:S2}(b) and \ref{fig:S2}(c) for $\varphi = 0^\circ$ and $45^\circ$, respectively.
Both patterns exhibit sharp streaks, indicating a smooth surface morphology and well-ordered in-plane crystallinity of the \ce{LaAg2Ge2} film.

Extended reciprocal-space mappings are presented in Figs.~\ref{fig:S2}(d) and \ref{fig:S2}(e).
At $\varphi = 0^\circ$, diffraction peaks corresponding to the \ce{LaAg2Ge2} (2\,0\,10) and (208) reflections are clearly observed along the $h = 2$ rod, together with the MgO (204) substrate peak.
This confirms that the $a$ and $b$ axes of the \ce{LaAg2Ge2} film are aligned with those of the MgO substrate, as illustrated in Fig.~\ref{fig:S2}(a).
At $\varphi = 45^\circ$, the \ce{LaAg2Ge2} (1\,1\,10) and (118) reflections appear along the $h = k = 1$ rod, confirming the fourfold in-plane symmetry of the film.
The relative intensities of the observed \ce{LaAg2Ge2} diffraction peaks are consistent with those calculated from the bulk crystal structure, indicating that the film preserves its intrinsic crystallographic symmetry.

\clearpage
\section{Origin of the dip/peak structure in the angle-dependent magnetoresistance ratio (ADMR)}

\begin{figure}[h]
\includegraphics[width=0.65\columnwidth]{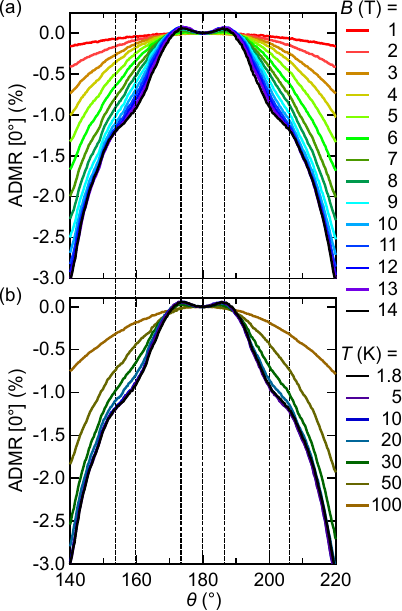}
\caption{
    Expanded view of the angle-dependent magnetoresistance ratio $\mathrm{ADMR} [0^\circ]\equiv \rho_{\mathrm{xx}}(\theta)/\rho_{\mathrm{xx}}(0^{\circ})-1$ in the vicinity of $\theta=180^{\circ}$.
    (a)~Field dependence at $T=1.8$~K. (b)~Temperature dependence at $B=14$~T.
    The angular positions of the features are independent of field and temperature.
}
\label{fig:S3}
\end{figure}

To clarify the origin of the dip-and-peak structures observed around $180^{\circ}$ in the ADMR of \ce{LaAg2Ge2} (Fig.~\ref{fig:S3}), we consider three mechanisms known to produce characteristic angular features in the magnetoresistance of metals and semimetals.

\subsection{Multiband ADMR from coexisting quasi-two-dimensional and three-dimensional Fermi-surface sheets}

In compensated multiband systems hosting Fermi-surface sheets of different dimensionality, the angle-dependent magnetoresistance can develop a complex profile with dips and peaks that reflect the distinct angular responses of each carrier pocket.
When the field is tilted to specific angles, the relative contributions of different sheets to the total conductivity change, and their competition can produce alternating dip-and-peak structures at angles governed by the Fermi-surface geometry.
Such behavior has been reported in the multiband systems \ce{ZrSiS}~\cite{Ali2016_Butterfly}, \ce{NbSb2} and \ce{TaSb2}~\cite{Yuan2016_Large}, TiB$_2$~\cite{Cheng2025_Anomalousc}, and \ce{W2As3}~\cite{Zhao2019_Extremely, Wang2019_Angledependentc}, where the interplay between quasi-two-dimensional and more three-dimensional pockets gives rise to nontrivial angular structures in $\rho_\mathrm{xx}(\theta)$.
Because \ce{LaAg2Ge2} is a compensated multiband system---as evidenced by the nonlinear Hall resistivity and the two-carrier analysis presented in the main text---a similar multiband origin for the dip-and-peak structures is plausible.
In this scenario, the angular positions of the features are determined by the Fermi-surface geometry and are therefore independent of both magnetic-field strength and temperature, consistent with our observations.

\subsection{Yamaji oscillations}

Yamaji oscillations are angular magnetoresistance oscillations (AMRO) characteristic of quasi-two-dimensional metals with a corrugated cylindrical Fermi surface~\cite{Yamaji1989_Angle}.
At specific tilt angles, all cross-sectional orbits on the corrugated cylinder share the same extremal area, producing peaks or dips in the interlayer resistivity $\rho_\mathrm{zz}$.
However, Yamaji oscillations are predominantly an interlayer ($\rho_\mathrm{zz}$) phenomenon, whereas the features in \ce{LaAg2Ge2} appear in the in-plane resistivity $\rho_\mathrm{xx}$.
Moreover, the Fermi wave vector estimated from the expected Yamaji periodicity, $k_\mathrm{F}=0.64$~\AA$^{-1}$ ($\sim$88\% of $k_a$), far exceeds the value consistent with the carrier density obtained from the Hall-effect analysis.
A Yamaji-oscillation origin is therefore unlikely.

\subsection{Lebed magic-angle resonances}

Lebed resonances produce sharp features in the angular magnetoresistance of quasi-one-dimensional conductors possessing open Fermi-surface sheets~\cite{Lebed1989_Theory, Osada1992_Resonance, Maki1992_Lebeda}.
At field orientations commensurate with the real-space lattice periodicity along the open-orbit direction, the interlayer conductivity is resonantly enhanced, generating dips in $\rho_\mathrm{zz}$.
Although $\arctan(a/c) = 21.6^{\circ}$ is close to the angular position of one of the observed dips in \ce{LaAg2Ge2}, band-structure calculations for Si- and Ge-based 122 compounds do not reveal quasi-one-dimensional open Fermi-surface sheets~\cite{Zwicknagl2007_Kondoa, DucDung2008_Haasb, Yasui2011_Electronica, Ciesielski2018_Electronic}, and the tetragonal crystal structure is incompatible with the low-dimensional topology required for Lebed resonances.
This mechanism is therefore unlikely as well.

\medskip
On the basis of the above analysis, we conclude that the dip-and-peak structures in the ADMR of \ce{LaAg2Ge2} most likely originate from its multiband electronic structure, in which the coexistence of Fermi-surface pockets with different dimensionalities and anisotropies produces a nontrivial angular profile in the in-plane magnetoresistance.
This interpretation is consistent with the field- and temperature-independent angular positions of both the dips and peaks, which are dictated by the intrinsic Fermi-surface geometry.

\end{document}